\documentclass[aps,pre,preprint,eqsecnum]{revtex4}

\usepackage{bm}
\usepackage{graphicx}  % standard LaTeX graphics tool
\usepackage{amsmath}   % useful for coding complex math
\def\Kp{{D}}
\def\bphi{\bm{\phi}}
\def\bpsi{{\bm{\psi}}}
\def\bomega{\bm{\omega}}
\def\bF{{\bf F}}
\def\bL{{\bf L}}
\def\bW{{\bf W}}
\def\eq{{\rm eq}}
\def\rmc{{\rm c}}
\def\rmd{{\rm d}}
\def\rme{{\rm e}}
\def\rmi{{\rm i}}
\def\rmw{{\rm w}}
\def\rmB{{\rm B}}
\def\rmI{{\rm I}}
\def\tw{t_\rmw}
\def\cH{{\cal H}}
\def\wcH{\widetilde\cH}
\def\total{{\rm tot}}
\def\extend{{\rm ext}}
\begin{document}
\title{Dynamical Gauge Theory for the $XY$ Gauge Glass Model}
\author{Yukiyasu Ozeki}
\affiliation{Department of Physics, Tokyo Institute of Technology, \\ 
Oh-okayama, Meguro-ku, Tokyo 152-8551, Japan}
\date{\today}
\begin{abstract}
Dynamical systems of the gauge glass are 
investigated by the method of the gauge transformation.
Both stochastic and deterministic dynamics are treated.
Several exact relations are derived among dynamical quantities 
such as equilibrium and nonequilibrium auto-correlation functions, 
relaxation functions of order parameter and internal energy.
They provide physical properties in terms of 
dynamics in the SG phase, a possible mixed phase and the Griffiths
phase, the multicritical dynamics and the aging phenomenon.
We also have a plausible argument for the absence of re-entrant 
transition in two or higher dimensions.
\end{abstract}
\maketitle
\section{Introduction}
The physics of disordered systems has been one of most fascinating
subjects for theorists and experimentalists.
The gauge glass is a typical example in this category 
and has attracted much attentions.
The model describes thermodynamics of various systems such as
disordered magnets with random Dzyaloshinskii-Moriya interaction
\cite{RubiSN83},
Josephson-junction arrays with positional disorder in a magnetic field
\cite{GranaK86},
vortex glasses
\cite{FishTY91},
and crystal systems on disordered substrates 
\cite{ChaFer95}. 
In three dimensions,
the spin glass (SG) transition 
\footnote{
This disordered phase is sometimes called ``the glass phase'' or 
``the vortex glass phase''.
However, in this paper, we call it the SG phase, since
they have the same properties.
}
for strongly disordered regime has been confirmed in the gauge glass systems
by numerical simulations 
\cite{HuseS90,RegeEA91,CiepBK91,LiNRS96,KosteS97,KosteA98,MaucoG98,
OlsonY00,AkinoK02} 
and renormalization group (RG) analyses \cite{Gingra91,CiepEA92},
which is consistent with experimental observations
\cite{KochEA89,GammSB91,DekkEK92}.

In two dimensions,
there is a controversy about the existence of glass-like phase
in strongly disordered regime.
Although the long-range SG order is denied in two dimensions
\cite{NishiK93}, 
there is a possibility of a quasi-long-range order
in which the SG correlation decays according to a power law
\cite{OzekiN93,Ozeki96}
like the ferromagnetic (FM) correlation in the Kosterlitz-Thouless (KT)
phase \cite{KosteT73,Koster74}.
Some numerical simulations supported the latter case
\cite{Li92,JeonKC95,KimEA97,ChoiP99},
while
experimental observation \cite{DekkEA92} 
and numerical simulations 
\cite{KosteS97,KosteA98,AkinoK02,RegerY93,HymaEA95} deny it.
In weakly disordered regime,
the instability of the KT phase against small
disorder is suggested by the perturbation expansion and 
the RG analysis \cite{Korshn93,MudryW99},
while it is denied by numerical simulations
\cite{ChaFer95,JeonKC95,KosteS97,MaucoG97}
and other RG analyses 
\cite{JeonKC95,NattEA95,KorshN96,Tang96,Scheid97}.
There is another controversy in two dimensions about 
the existence of re-entrant transition from the KT phase to the non-KT one.
Earlier works with real space RG analyses
suggest re-entrance \cite{RubiSN83,GranaK86,ForrEA88,PaczuK91},
however
the analysis has been modified and provides the absence of it
\cite{NattEA95,KorshN96,Tang96,Scheid97}. 
The same results are obtained by
Monte Carlo simulations
\cite{ChakrD88,ForrBL90,ChaFer95,JeonKC95,KosteS97,MaucoG98} 
and the RG analyses
\cite{GingrS92,NattEA95}. 

Since randomness and frustration make it difficult to examine 
short-range systems analytically as well as numerically,
only few things have been confirmed definitely
in spite of huge number of studies.
Thus, it is highly desirable to have analytic method 
for understanding the thermodynamic and critical behaviors 
in random spin systems.
The method of gauge transformation 
is a useful technique to derive analytic results 
in gauge symmetric random spin systems
\cite{Nishim81,Kitata92,OzekiN93,Ozeki95,Ozeki96,Ozeki96a,Ozeki97,OzekiI98}.
It was firstly applied to the Ising SG models \cite{Nishim81},
and provides the internal energy and an upper bound on the 
specific heat exactly as non-singular functions of the temperature 
on the Nishimori line in the temperature-randomness phase diagram. 
Further, it is proven that the boundary between the FM
and non-FM phases
in the temperature lower than the multicritical point 
(MCP, see Fig.\ \ref{PhaseD})
at which paramagnetic (PM), FM and SG phases meet and
the Nishimori line intersects, is vertical or re-entrant. 
Using a model with a modified probability distribution of randomness,
the verticality is concluded \cite{Kitata92},
which implies the absence of re-entrance:
Although the theory needs a plausible but unproved assumption 
on the thermodynamic behavior of the modified model, 
the result is consistent with other theories 
and experiments of Ising-like spin glasses \cite{BindeY86}.
The gauge transformation has been applied 
to gauge glasses with various symmetries \cite{OzekiN93}
including the $U(1)$ symmetry, {\it i.e.} the $XY$ gauge glass model.
The same results as in the Ising case are derived.
The absence of re-entrance is concluded for any dimensional model
with a plausible assumption for the modified model,
if the FM (or KT) phase is stable in disordered regime.
The gauge transformation is also extended to dynamical systems
\cite{Ozeki95,Ozeki96a,Ozeki97,OzekiI98}.
For the Ising SG case, some important relations are 
derived for nonequilibrium auto-correlation functions, order parameter
and energy relaxations.
They provide discussions on the absence of re-entrance,
the regime of the mixed phase in the phase diagram, 
the multi-critical dynamics, and the equivalence of two nonequilibrium 
processes in the SG phase.

The dynamics is one of most important aspects in disordered systems
because of the slow relaxation \cite{BindeY86,FischH91}.
In the case of standard SG theory,
dynamical properties have been investigated 
for the mean field model \cite{SompoZ81}, 
for the Griffiths phase in short-range systems
\cite{RandSP85,Ogiels85,TakanM95,KomoHT95},
for the aging phenomenon
\cite{KoperH88,FisheH88,SibanH89,Rieger93,CugliK94} 
and so on.
The aging phenomenon \cite{Struik78} is a typical realization of slow
dynamics, especially for SG materials 
\cite{LundEA83,NordEA87,VincHO92,LeflEA94}.
In contrast to the standard SG systems, not many
studies on the dynamical properties have been performed for gauge glass
systems.
It would be helpful for future investigations to present exact results
including dynamical properties.

In this paper, we study the gauge transformation for dynamical 
systems of the gauge glass model.
In the static case, the gauge symmetric models can be treated in the
same manner as the Ising SG \cite{OzekiN93}.
It is also straightforward to apply the dynamical gauge theory for
the Ising SG \cite{Ozeki95,Ozeki97,OzekiI98},
in which the stochastic dynamics is treated, to the gauge glass model.
Since the gauge glass system has a continuous spin space,
one can consider a deterministic dynamics with equations of motion,
which is also treated in the present theory.
Several exact relations are derived among dynamical quantities 
such as equilibrium and nonequilibrium auto-correlation functions, 
relaxation functions of order parameter and internal energy.
Using them, we discuss physical properties including
the aging, the multicritical dynamics, 
dynamics in the SG, a possible mixed phase and Griffiths phase,
and the absence of re-entrant transition in two or higher dimensions.

The organization of the paper is as follows.
In section \ref{sec:Model}, the gauge glass model associated with the plane
rotator is introduced. The probability distribution is chosen so that 
the system has the gauge symmetry.
The results of the static gauge theory is summarized.
In section \ref{sec:SD},
a stochastic dynamics is introduced using the master equation
formalism, and some dynamical quantities are defined. 
The gauge transformation is introduced for
this system.
Some exact relations are derived for dynamical quantities.
In section \ref{sec:PP}, physical properties are discussed 
by use of the exact relations obtained.
In section \ref{sec:HD}, equations of motion are
introduced for the same gauge symmetric system, and 
the gauge transformation is applied.
It is shown that the same relations can be derived 
for several choices of dynamics.
The last section is devoted to the summary.
\section{Gauge Symmetric Gauge Glass Model}
\label{sec:Model}
The Hamiltonian we consider is
\begin{equation}
\cH ~=~
J\wcH(\bphi,\bomega) ~=~
-J\,\sum_{\langle ij \rangle}\, \cos(\phi_i -\phi_j +\omega_{ij} ) ,
\label{defH}
\end{equation}
where $\phi_i ,\omega_{ij} \in [-\pi,\pi )$.
The set $\bphi =(\phi_1 ,\phi_2 ,\cdots ,\phi_N )$ 
represents a configuration of total $N$ spins, and
$\bomega$ represents a configuration of total $N_\rmB$ 
quenched random variables associated to each bond.
The summation of $\langle ij\rangle$ is taken over all bonds; 
we make no restrictions on the type or the dimensions 
of lattice, whereas one may suppose a usual nearest-neighbor 
interaction on the $d$-dimensional hypercubic lattice.
The variable $\omega_{ij}$ has an odd property;
$\omega_{ji} =-\omega_{ij}$ 
\footnote{
The sign of $\omega_{ij}$ in eq.\ (\ref{defH}) is not important,
since the distribution of $\omega_{ij}$ (\ref{defP}) is an even
function.}. 
For a particular bond configuration, the thermal equilibrium
distribution at temperature $T=J/k_\rmB K$ is given by 
\begin{equation}
\rho_\eq (\bphi; K,\bomega)~=~
{\exp\left(-K\wcH (\bphi;\bomega)\right)
\over
Z(K;\bomega )}.
\label{rhoeq}
\end{equation}
The partition function is defined by
\begin{equation}
Z(K;\bomega )~=~\int\rmd\bphi~
\exp\left(-K\wcH (\bphi;\bomega)\right),
\end{equation}
where the integration of $\rmd\bphi$ denotes 
the multiple integrations for the $N$ spin variables $\phi_i$.
The probability distribution for randomness is taken as
\begin{equation}
P(\bomega ;\Kp )~=~
\prod_{\langle ij\rangle}\,
{\exp (\Kp \cos\omega_{ij} )\over 2\pi I_0 (\Kp )} ~=~
{\exp\left(-\Kp \wcH (\bF;\bomega )\right)
\over
Y(\Kp )},
\label{defP}
\end{equation}
where 
$\bF =(0,0,\cdots ,0)$ represents an all-aligned spin state,
$I_n(\Kp)$
is the modified Bessel function, and 
\begin{equation}
Y(\Kp )\equiv \big(2\pi I_0 (\Kp )\big)^{N_\rmB}.
\end{equation}
In this distribution, which is central to the present paper 
(see also Fig.\ \ref{Pomega} below), 
the variable $\Kp$ controls the strength of
randomness; $\Kp =0$ and $\infty$ correspond to 
the full random case and the non-random case, respectively.
The random average is denoted by
\begin{equation}
[\cdots ]_\rmc ~=~
\int \rmd\bomega~P(\bomega ;\Kp )~\cdots~.
\label{rAV}
\end{equation}
The integration of $\rmd\bomega$ denotes 
the multiple integrations for the $N_B$ random variables $\omega_{ij}$.
In the following, for simplicity, we will sometimes omit the dependence 
on spin set $\bphi$, bond set $\bomega$, 
inverse temperature $K$ and/or randomness $\Kp$ 
from functions defined above, if they are trivial or unimportant.

The gauge transformations for functions of 
$\bphi$ and $\bomega$ are defined by
\begin{eqnarray}
U_\bpsi~&:&~\phi_i     ~\longrightarrow~\phi_i      -\psi_i 
\\
V_\bpsi~&:&~\omega_{ij}~\longrightarrow~\omega_{ij} +\psi_i -\psi_j,
\end{eqnarray}
where $\bpsi =(\psi_1 ,\psi_2 ,\cdots ,\psi_N)$ 
is an arbitrary state of $N$ spins.
Each set of transformation forms a group.
While variables $\phi_i$ and $\omega_{ij}$ possibly take
values out of the domain, $[-\pi,\pi)$, in this transformation,
we always identify them with a residue of $2\pi$.
The Hamiltonian (\ref{defH})
is invariant under the transformation $U_\bpsi V_\bpsi$:
\begin{equation}
U_\bpsi V_\bpsi~
\wcH(\bphi;\bomega )~=~
\wcH(\bphi;\bomega ).
\label{giH}
\end{equation}
The gauge transformation of the distribution (\ref{defP}) is given by
\begin{equation}
V_\bpsi ~P(\bomega ;\Kp ) ~=~
{\exp\left(-\Kp \wcH (\bpsi;\bomega )\right)
\over
Y(\Kp )}~=~
{Z(\Kp;\bomega)
\over
Y(\Kp)}
~\rho_\eq (\bpsi;\Kp,\bomega),
\label{gtP}
\end{equation}
where
we note that the spin set in $\wcH$ and the temperature are 
different from the usual ones.
Another important property in the theory is the invariant integral 
for $\bphi$ and $\bomega$:
\begin{eqnarray}
\int \rmd\bphi~\cdots ~&=&~
\int \rmd\bphi~U_\bpsi~\cdots ,
\label{iiPU}\\
\int \rmd\bomega~ \cdots ~&=&~
\int \rmd\bomega~ V_\bpsi~ \cdots 
\label{iiOV}
\end{eqnarray} 
Using eqs.\ (\ref{giH}) and (\ref{gtP}), 
one can show \cite{OzekiN93} that 
the partition function is gauge invariant under $V_\bpsi$,
\begin{equation}
V_\bpsi~Z(K;\bomega) ~=~Z(K;\bomega),
\label{giZ}
\end{equation}
providing the invariance of thermal distribution, 
\begin{equation}
U_\bpsi\,V_\bpsi~\rho_\eq(\bphi;K,\bomega) ~=~\rho_\eq(\bphi;K,\bomega). 
\label{girhoeq}
\end{equation}
Then, we obtain the gauge invariance of the averaged energy;
\begin{equation}
V_\bpsi~ \langle\wcH\rangle_K~=~\langle\wcH\rangle_K.
\label{giAH}
\end{equation}
Note that we will use the terminology "gauge invariant" only for functions of 
${\bomega}$ invariant under $V_\bpsi$.
Since the lhs of eq.\ (\ref{iiOV}) is independent of $\bpsi$,
it is convenient to derive another expression of random average 
instead of eq.\ (\ref{rAV});
\begin{equation}
[\cdots]_\rmc~=~
\int \rmd\bomega\,\rmd\bpsi~
{Z(\Kp;\bomega )\over (2\pi )^N Y(\Kp )}~ 
\rho_\eq (\bpsi;\Kp,\bomega)~
V_\bpsi~ \cdots .
\label{rAVgt}
\end{equation}
The random average of gauge 
invariant function $Q(\bomega )$ can be expressed as
\begin{equation}
\left[ Q(\bomega )\right]_\rmc~=~
\int \rmd\bomega~ 
{Z(\Kp;\bomega )\over (2\pi )^N Y(\Kp )}~Q(\bomega ). 
\label{rAVgi}
\end{equation}

\begin{figure}
\includegraphics[width=0.5\textwidth]{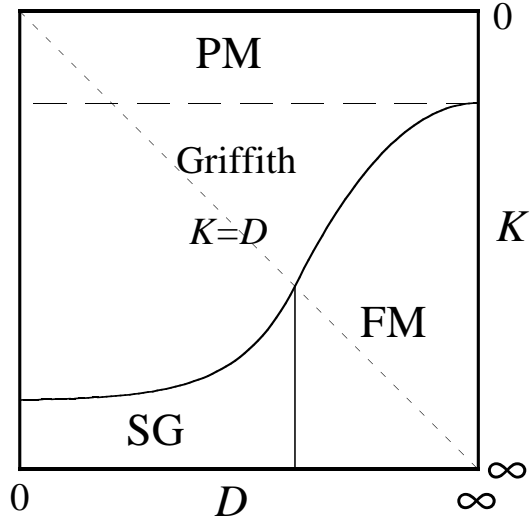}
\caption{Typical phase diagram of gauge glass model 
in the $\Kp$-$K$ plane ($\Kp$ controls the randomness 
and $K=J/k_\rmB T$).
The dashed line $K=\Kp$ is the Nishimori line.
The FM phase boundary in low temperature region
is assumed to be vertical.
A possible regime of the Griffiths phase is indicated.
\label{PhaseD}}
\end{figure}

In three or higher dimensions, 
the topology of the $K$-$\Kp$ phase diagram is expected as 
in Fig.\ \ref{PhaseD}, 
where PM, FM and SG phases appear.
A possible Griffiths phase is also indicated.
In two dimensions, the KT phase appears instead of the FM phase.
The SG phase disappears or becomes a possible KT-like SG phase.
The dashed line $K=\Kp$ is called the Nishimori-line
around which following properties have been found
from the static gauge theory
\cite{OzekiN93,Ozeki96a}: 
\begin{itemize}
\item
The energy and the upper bound of the specific heat are expressed 
by analytic functions of temperature;
\begin{eqnarray}
\label{eq:E}
E(\Kp,\Kp) &=& -J{\rmd\over\rmd\Kp} \ln Y(\Kp) 
=-N_\rmB J{I_1(\Kp)\over I_0(\Kp)},\\
C(\Kp,\Kp) &\le& k_\rmB \Kp^2 {\rmd^2\over\rmd\Kp^2} \ln Y(\Kp).
\end{eqnarray}
\item
The FM and SG order parameters coincide with each other;
\begin{equation}
m(\Kp,\Kp)=q(\Kp,\Kp).
\label{eq:mq}
\end{equation}
Thus, no SG phase appears on this line, since $m=0$ in the SG phase.
\item
The line is likely to intersect the MCP of PM, FM and SG phases.
\item
The phase boundary between the FM (or KT) and the SG (or PM) phases 
below this line must be vertical to the $\Kp$-axis
or re-entrant
%, not smelling to small $\Kp$ side
\cite{OzekiN93}.
A plausible argnument can be made for the verticality.
\end{itemize}

\begin{figure}
\includegraphics[width=0.6\textwidth]{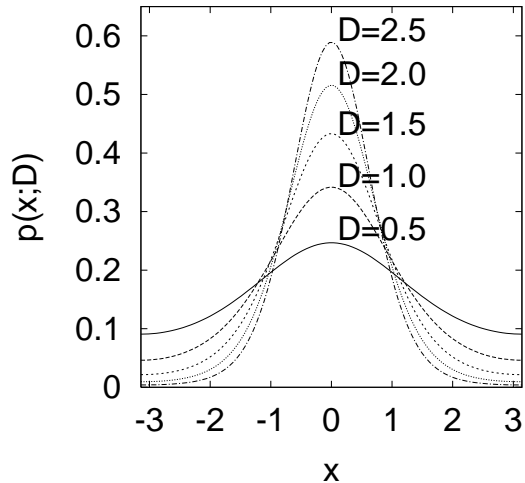}
\caption{The probability distribution function for one bond
$p(x;\Kp) \equiv \exp(\Kp\cos x)/(2\pi I_0(\Kp))$.}
\label{Pomega}
\end{figure}

It is remarked that the probability distribution function (\ref{defP})
is different from the usual Gaussian form, which has been studied 
in many cases.
This is because that we need a gauge symmetric form
in the theory, 
which restricts the distribution function relating to the Hamiltonian
as in eq.\ (\ref{defP}).
It is shown that 
the same properties can also be proven for another gauge
symmetric $U(1)$ model, the Villain model \cite{Villai75}
with Gaussian random gauge,
in which the local Boltzmann factor is given by
\begin{equation}
\rme^{V(\phi_i-\phi_j+\omega_{ij})}
=\sum_{n=-\infty}^{\infty}
\rme^{-K(\phi_i-\phi_j+\omega_{ij}-2\pi n)^2/2}.
\end{equation}
We consider that physical properties are equivalent in 
these gauge glass models irrespective of the detail of distribution $P$.
In fact, as seen in Fig.\ \ref{Pomega}, the shape of the function 
(\ref{defP}) is quite similar to the Gaussian.
As will be seen later, the present theory can be applied to the Villain
model in the case of the stochastic dynamics in section \ref{sec:SD}.
However, we proceed our present theory by the rotator model.
It is somehow difficult to apply in the case of the deterministic
dynamics, since the explicit form of the Hamiltonian, which is necessary
in the equations of motion, is not unique. 

\section{Stochastic Dynamics}
\label{sec:SD}
First, we consider the stochastic dynamics following the
results of Ising SG \cite{Ozeki95,Ozeki97,OzekiI98}.

\subsection{Master Equation}
We consider a Markov process for a fixed bond configuration,
in which the density of state evolves with the master
equation \cite{VanKam81},
\begin{equation}
{\rmd\over\rmd t} \rho_t (\bphi )~=~
\int \rmd\bphi'~\left\{
W(\bphi| \bphi')\, \rho_t (\bphi' )-
W(\bphi'| \bphi)\, \rho_t (\bphi )\right\}.
\label{eqMaster}
\end{equation}
The transition probability $W(\bphi |\bphi')$ 
is non-negative and satisfies the condition of the detailed balance,
\begin{equation}
W(\bphi |\bphi')\, \rho_\eq (\bphi';K,\bomega )~=~
W(\bphi' |\bphi)\, \rho_\eq (\bphi;K,\bomega ), 
\label{eqDV}
\end{equation}
which guarantees the stability of the equilibrium distribution 
$\rho_\eq$.
The condition (\ref{eqDV}) is automatically satisfied 
if one uses the expression with a symmetric matrix $w(\bphi |\bphi')$:
\begin{equation}
W(\bphi|\bphi')\, =
\,{w(\bphi|\bphi')\over\rho_\eq (\bphi' )}.
\end{equation}
The function $w(\bphi|\bphi')$ 
depends on the detail of dynamics considered.
As in \cite{Ozeki95}, I assume
\begin{equation}
w(\bphi|\bphi') =w_0\,
\delta_1 [\bphi,\bphi' ]~
\rho_\eq 
(\bphi  )^{\Theta (\Delta [\bphi,\bphi' ])}
\rho_\eq 
(\bphi' )^{\Theta (\Delta [\bphi' ,\bphi ])}
\label{wMet}
\end{equation}
for the single-spin-flip Metropolis dynamics \cite{MeRRTT53}, and 
\begin{equation}
w(\bphi|\bphi') = w_0\,\sum_i
{\delta_{1i\,} [\bphi,\bphi' ]\,
\rho_\eq (\bphi  )\,\rho_\eq (\bphi' )\over
\int \rmd\bphi''\, \delta_{1i\,} [\bphi'',\bphi' ]\,
\rho_\eq (\bphi'' )}
\label{wHB}
\end{equation}
for the single-spin-flip heat-bath dynamics, 
where $\Theta(x)$ denotes the Heaviside function,
\begin{equation}
\Delta [{\bphi,\bphi'} ]\equiv 
\wcH (\bphi ;\bomega)-
\wcH (\bphi';\bomega)
\end{equation}
is the energy difference of two states,
\begin{eqnarray}
\delta_1 [\bphi,\bphi' ]~&\equiv&~
\sum_i \delta_{1i\,} [\bphi,\bphi' ]\\
\delta_{1i\,} [\bphi,\bphi' ]~&\equiv&~
\prod_{j\ne i} \delta (\phi_j -\phi'_j )
\end{eqnarray} 
are the single-spin-flip and the $i$th-spin-flip operators, respectively. 
Note that the denominator of the summand in eq.\ (\ref{wHB}) is unchanged
if the variable $\bphi'$ in the argument of 
$\delta_{1i\,} [\bphi'',\bphi' ]$
is replaced by $\bphi$ providing the symmetry of 
$w(\bphi|\bphi')$.

Since the diagonal element plays no role both 
in eqs.\ (\ref{eqMaster}) and (\ref{eqDV}), 
one may define it in such a way as
\begin{equation}
W(\bphi|\bphi')\, =
\,{w(\bphi|\bphi')\over\rho_\eq (\bphi' )}
- \delta (\bphi-\bphi') \int \rmd\bphi''~
{w(\bphi'' |\bphi)\over\rho_\eq (\bphi )} .
\end{equation}
Then, the master equation (\ref{eqMaster}) is reduced to
\begin{equation}
{\rmd\over\rmd t} \rho_t (\bphi )~=~
\int \rmd\bphi'~
W(\bphi| \bphi')~\rho_t (\bphi' ),
\end{equation}
and the solution is given by 
\begin{equation}
\rho_t (\bphi )=\int \rmd\bphi'~
\left\langle\bphi \left|\rme^{t\bW}\right|\bphi' 
\right\rangle\,\rho_0 (\bphi' ).
\end{equation}
The function
$\left\langle\bphi \left|\rme^{t\bW}\right|\bphi' 
\right\rangle$
is the integration kernel for time-evolution defined by
\begin{eqnarray}
\left\langle\bphi \left|\rme^{t\bW}\right|\bphi' 
\right\rangle~&\equiv&~
\sum_{n=0}^\infty {t^n \over n!}
\left\langle\bphi \left| \bW^{n}\right|\bphi' 
\right\rangle, \\
\left\langle\bphi \left| \bW^{n}\right|\bphi' \right\rangle 
~&\equiv&~
\int \left(
\prod_{\ell =1}^{n-1} \rmd\bphi^{(\ell )} %\cdot
\right)~
\prod_{k=0}^{n-1} W(\bphi^{(k+1)} |\bphi^{(k)}),
\label{defETW}
\end{eqnarray}
where $\bphi^{(n)} =\bphi$ and $\bphi^{(0)} =\bphi'$.
It has been proven that all eigenvalues of the operator $\bW$ 
with (\ref{eqDV}) 
are real and negative except the zero eigenvalue corresponding 
to the equilibrium distribution \cite{VanKam81}.
Thus, all solutions with any initial 
conditions tend to the equilibrium distribution as $t\to\infty$.

\subsection{Dynamical Processes}
\begin{figure}
\includegraphics[width=0.46\textwidth]{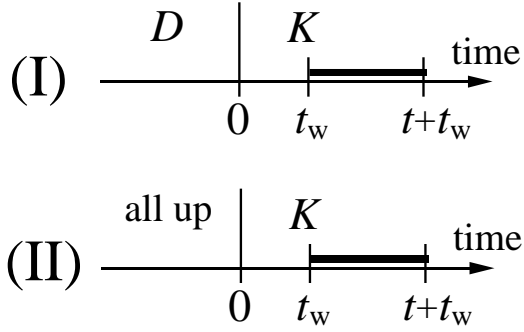}
\caption{
Illustrations of two nonequilibrium processes, I and II.
Between $t_{\rm w}$ and $t+t_{\rm w}$, the correlation is measured.
\label{process}}
\end{figure}
We examine two nonequilibrium processes I and II in Fig.\ \ref{process}.
The process I is related to the zero-field-cooling (ZFC).
At the initial time $t=0$, 
the system is kept in the equilibrium state 
with a temperature $\Kp$ and zero field;
hereafter, we use a terminology ``temperature" for $K$ and $\Kp$ 
instead of ``inverse temperature". 
The temperature is immediately changed
(usually quenched) and the system relaxes 
in a heat bath with another temperature $K$ in $t>0$.
The average for dynamical ensembles in this process 
is denoted by $\langle\cdots\rangle_K^{\Kp}$.
The process II is related to the field quench (FQ) 
\cite{NordEA87,Rieger93}.
The system starts from an all-aligned state ${\bf F}$ at $t=0$ and 
relaxes in the same heat bath as in the process I in $t>0$;
the average is denoted by $\langle\cdots\rangle_K^\bF$.
Since the all-aligned state provides the strong-field limit,
this represents the process with
a field quenched from $\infty$ to zero at $t=0$.
Note that the FQ is not equivalent to the field cooling 
in which the applied field is weaker
and is quenched after the waiting.

We define
the waiting-time dependence of nonequilibrium auto-correlation functions
\cite{Rieger93,CugliK94}
for the processes I and II.
In the language of master equation, that for the process I is 
expressed as
\begin{eqnarray}
\big\langle \rme^{\rmi\phi_i (t+\tw) -\rmi\phi_i (\tw)} 
\big\rangle_K^\Kp =
\int \rmd\bphi\,\rmd\bphi'\,\rmd\bphi_0~
\left\langle\bphi \left|\rme^{t\bW}\right|\bphi'  \right\rangle 
\left\langle\bphi' \left|\rme^{\tw\bW}\right|\bphi_0\right\rangle \,
\rho_\eq (\bphi_0;\Kp )~
\rme^{\rmi\phi_i -\rmi\phi'_i}.
\nonumber\\ 
\label{AC1}
\end{eqnarray}
When $K=\Kp$, the average $\langle\cdots\rangle_\Kp^\Kp$
represents the dynamical average in the equilibrium process
$\langle\cdots\rangle_\Kp^\eq$.
In this case, the auto-correlation function (\ref{AC1}) becomes 
independent of $\tw$ since the thermal distribution 
$\rho_\eq (\phi;K,\bomega)$ is an eigenfunction of 
$W(\bphi'|\bphi)$, {\it i.e.}
\begin{eqnarray}
\label{ACeq}
\big\langle \rme^{\rmi\phi_i (t+\tw) -\rmi\phi_i (\tw)} 
\big\rangle_\Kp^\Kp &=&
\big\langle \rme^{\rmi\phi_i (t) -\rmi\phi_i (0)} 
\big\rangle_\Kp^\eq\\
\nonumber 
&=&
\int \rmd\bphi\,\rmd\bphi'~
\left\langle\bphi \left|\rme^{t\bW}\right|\bphi' \right\rangle\, 
\rho_\eq (\bphi';K )\,
\rme^{\rmi\phi_i -\rmi\phi'_i} .
\end{eqnarray}
If one defines an auto-correlation function for the process relaxed 
from a fixed state $\bpsi$ 
--- let us call this process the process II' ---
\begin{equation}
\big\langle \rme^{\rmi\phi_i (t+\tw) -\rmi\phi_i (\tw)} 
\big\rangle_K^\bpsi = 
\int \rmd\bphi\,\rmd\bphi'~
\left\langle\bphi \left|\rme^{t\bW}\right|\bphi'  \right\rangle 
\left\langle\bphi' \left|\rme^{\tw\bW}\right|\bpsi\right\rangle \,
\rme^{\rmi\phi_i -\rmi\phi'_i},
\label{ACpsi}
\end{equation}
that for the process II is expressed as
$\big\langle \rme^{\rmi\phi_i (t+\tw) -\rmi\phi_i (\tw)} 
\big\rangle_K^\bF$, and
eq.\ (\ref{AC1}) is rewritten as
\begin{equation}
\big\langle \rme^{\rmi\phi_i (t+\tw) -\rmi\phi_i (\tw)} 
\big\rangle_K^\Kp =
\int \rmd\bpsi~
\rho_\eq (\bpsi;\Kp )~
\big\langle \rme^{\rmi\phi_i (t+\tw) -\rmi\phi_i (\tw)} 
\big\rangle_K^\bpsi. 
\label{AC1r}
\end{equation}
In the long-time limit $(t\to\infty )$, the thermodynamic limit of 
$\left[\big\langle \rme^{\rmi\phi_i (t) -\rmi\phi_i (0)} 
\big\rangle_K^\eq \right]_\rmc$ 
converges to the SG order parameter $q_{\rm EA}$,
which is non-vanishing in FM or SG
phases \cite{BindeY86,FischH91}. 
If one sets $\tw=0$, equation (\ref{ACpsi}) with $\bpsi=\bF$ becomes
\begin{equation}
\big\langle \rme^{\rmi\phi_i (t)} 
\big\rangle_K^\bF =
\int \rmd\bphi \,
\left\langle\bphi \left|\rme^{t\bW}\right|\bF 
\right\rangle \, 
\rme^{\rmi\phi_i } ,
\end{equation}
which is the FM order parameter at time $t$ relaxed
from the complete FM state, $\bF$, at $t=0$.
When $t\to\infty$, the thermodynamic limit of 
$\left[\big\langle \rme^{\rmi\phi_i (t)} 
\big\rangle_K^\bF \right]_\rmc$ 
approaches the spontaneous ordering.
Note that all quantities defined above containing the exponential
$\rme^{\rmi\phi_i}$ have real values in the average
because of the spin-flip symmetry, $\phi_i \to \phi_i +\pi$,
for the Hamiltonian and the matrix $W(\bphi |\bphi')$.

Another relaxation functions, the nonequilibrium relaxations 
of exchange energy, are defined as 
\begin{eqnarray}
\big\langle \cH (t) \big\rangle_K^\Kp~&=&~
J\int \rmd\bphi \,\rmd\bphi'~
\left\langle\bphi \left|\rme^{t\bW}\right|\bphi' 
\right\rangle\,
\rho_\eq(\bphi';\Kp,\bomega)\,
\wcH (\bphi;\bomega)
\label{dynHD}\\
\big\langle \cH (t) \big\rangle_K^\bF~&=&~
J\int \rmd\bphi \,
\left\langle\bphi \left|\rme^{t\bW}\right|\bF 
\right\rangle\,
\wcH (\bphi;\bomega). 
\label{dynHF}
\end{eqnarray}
\subsection{Gauge Transformation}
Let us examine the gauge transformation of the functions defined above. 
We introduce the transformation for $\bphi'$,
\begin{equation}
U'_\bpsi ~:~ \phi'_i ~\longrightarrow~ \phi'_i -\psi_i, 
\end{equation}
to show the invariance 
\begin{equation}
U_\bpsi U'_\bpsi\, V_\bpsi~
\left\langle\bphi \left|\rme^{t\bW}\right|\bphi' 
\right\rangle ~=~
\left\langle\bphi \left|\rme^{t\bW}\right|\bphi' 
\right\rangle .
\label{giETW}
\end{equation}
{}From eqs.\ (\ref{giH})-(\ref{giZ}), it is easy to see that 
$\rho_\eq(\bphi;K,\bomega)$,
$\delta_{1i\,} [\bphi ,\bphi' ]$ and 
$\Delta [\bphi ,\bphi' ]$ 
are invariant under 
$U_\bpsi U'_\bpsi\, V_\bpsi$.
This yields the invariance
\begin{equation}
U_\bpsi U'_\bpsi\, V_\bpsi~
w(\bphi|\bphi')~=~
w(\bphi|\bphi')
\label{gilW}
\end{equation}
in the case of the Metropolis (\ref{wMet}) and 
the heat-bath dynamics (\ref{wHB}).
It is natural to restrict the dynamics in such a way as 
eq.\ (\ref{gilW}) holds.
This yields the invariance,
\begin{equation}
U_\bpsi U'_\bpsi\, V_\bpsi~
W(\bphi|\bphi')~=~
W(\bphi|\bphi') ,
\label{giW}
\end{equation}
where the invariant integral for $\bphi''$ like eq.\ (\ref{iiPU}) 
is necessary to derive for the case $\bphi =\bphi'$. 
Let us consider term by term in the expansion of 
$\left\langle\bphi \left|\rme^{t\bW}\right|\bphi' 
\right\rangle $, eq. (\ref{defETW}).
Similarly to $\bphi$ and $\bphi'$, 
the gauge transformations for $\bphi^{(k)}$ $(k=1,2,\cdots,n-1 )$
are introduced as
\begin{equation}
U^{(k)}_\bpsi ~:~ \phi_i^{(k)} 
~\longrightarrow~ \phi_i^{(k)} -\psi_i.
\end{equation}
Then, equation (\ref{giW}) with the invariant integral for $\bphi^{(k)}$ 
like (\ref{iiPU}) leads to
\begin{eqnarray}
U_\bpsi U'_\bpsi\, V_\bpsi~
\left\langle\bphi \left| \bW^{n}\right|\bphi' 
\right\rangle 
{}&=&~ 
\int \left(
\prod_{\ell =1}^{n-1} \rmd\bphi^{(\ell )}\,
U^{(\ell )}_\bpsi
\right)~ 
U_\bpsi U'_\bpsi~ V_\bpsi~
\prod_{k=0}^{n-1} W(\bphi^{(k+1)} |\bphi^{(k)}) 
\nonumber \\ 
{}&=&~ 
\int \left(
\prod_{\ell =1}^{n-1} \rmd\bphi^{(\ell )}
\right)~
\prod_{k=0}^{n-1} W(\bphi^{(k+1)} |\bphi^{(k)}),
\end{eqnarray}
providing the invariance (\ref{giETW}).

{}From eqs.\ (\ref{girhoeq}) and (\ref{giETW}) with the invariant 
integral (\ref{iiPU}) taken into account, one can show that
the auto-correlation function for the process I
is gauge invariant as
\begin{eqnarray}
\label{giACeq}
V_\bpsi~
\big\langle \rme^{\rmi\phi_i (t) -\rmi\phi_i (0)} 
\big\rangle_K^\eq ~&=&~
\big\langle \rme^{\rmi\phi_i (t) -\rmi\phi_i (0)} 
\big\rangle_K^\eq,\\
V_\bpsi~
\big\langle \rme^{\rmi\phi_i (t+\tw) -\rmi\phi_i (\tw)} 
\big\rangle_K^\Kp ~&=&~
\big\langle \rme^{\rmi\phi_i (t+\tw) -\rmi\phi_i (\tw)} 
\big\rangle_K^\Kp,
\label{giAC1}
\end{eqnarray}
while that for the process II is transformed as
\begin{eqnarray}
\nonumber 
V_\bpsi~
\big\langle \rme^{\rmi\phi_i (t+\tw) -\rmi\phi_i (\tw)} 
\big\rangle_K^\bF~&=&~
\int \rmd\bphi\,\rmd\bphi'~ U_\bpsi U'_\bpsi V_\bpsi 
\left\langle\bphi \left|\rme^{t\bW}\right|\bphi'  \right\rangle 
\left\langle\bphi' \left|\rme^{\tw\bW}\right|\bF\right\rangle \,
\rme^{\rmi\phi_i -\rmi\phi'_i}
\\
{}&=&~
\big\langle \rme^{\rmi\phi_i (t+\tw) -\rmi\phi_i (\tw)} 
\big\rangle_K^\bpsi
\label{gtAC2}
\end{eqnarray}
where we use the invariant integral (\ref{iiPU}), and
\begin{equation}
U_\bpsi V_\bpsi\, 
\left\langle\bphi \left|\rme^{t\bW}
\right| \bphi' \right\rangle ~=~
U'_{-\bpsi}  
\left\langle\bphi \left|\rme^{t\bW}
\right| \bphi' \right\rangle ~=~
\left\langle\bphi \left|\rme^{t\bW}
\right| \bphi'+\bpsi \right\rangle
\end{equation}
obtained from eq.\ (\ref{giETW}).
The variables $-\bpsi$ and $\bphi' +\bpsi$
represent the spin states, 
$(-\psi_1 ,-\psi_2 ,\cdots ,-\psi_N )$ and
$(\phi'_1 +\psi_1 ,\phi'_2 +\psi_2 ,\cdots ,\phi'_N +\psi_N )$,
respectively.
Using eqs.\ (\ref{rAVgt}), (\ref{AC1r}) and (\ref{gtAC2}),
we derive  
\begin{equation}
\left[
\big\langle \rme^{\rmi\phi_i (t+\tw) -\rmi\phi_i (\tw)} 
\big\rangle_K^\bF 
\right]_\rmc ~=~
\int \rmd\bomega~
{Z(\Kp;\bomega )\over (2\pi )^N Y(\Kp )}~ 
\big\langle \rme^{\rmi\phi_i (t+\tw) -\rmi\phi_i (\tw)} 
\big\rangle_K^\Kp .
\label{cavAC2}
\end{equation}
Because of the gauge invariance (\ref{giAC1}), 
equation (\ref{cavAC2}) is rewritten by eq.\ (\ref{rAVgi}) as
\begin{equation}
\left[
\big\langle \rme^{\rmi\phi_i (t+\tw) -\rmi\phi_i (\tw)} 
\big\rangle_K^\bF \right]_\rmc 
~=~
\left[\big\langle \rme^{\rmi\phi_i (t+\tw) -\rmi\phi_i (\tw)} 
\big\rangle_K^\Kp \right]_\rmc 
\label{AR} 
\end{equation}
for any waiting time $t_{\rm w}$, any time interval $t$,
any temperature $K$ and any degree of randomness $\Kp$.
We call it the ``aging relation'', since 
it relates aging phenomena in two distinct processes
whatever waiting-time dependence is essential for the aging.
{}From the invariance (\ref{giH}), similar relation can be derived 
for the energy defined in eqs.\ (\ref{dynHD}) and (\ref{dynHF}):
\begin{equation}
\left[\big\langle \cH (t) 
\big\rangle_K^\bF \right]_\rmc~=~
\left[\big\langle \cH (t) 
\big\rangle_K^\Kp \right]_\rmc .
\label{jHH}
\end{equation}
\section{Physical Properties}
\label{sec:PP}
The aging relation (\ref{AR}) and eq.\ (\ref{jHH}) include
parameters $K$ and $\Kp$ characterizing the relaxation in $t>0$.
For each phase, one can examine the physical meaning of the relation
by choosing $(K,\Kp)$ appropriately.
Note that the initial temperature of the process I is always located 
on $K=\Kp$.
\subsection{Dynamics in the SG Phase}
First, let us consider the region with 
sufficiently strong randomness (small $\Kp$) 
and sufficiently low temperatures (large $K$), 
where the SG phase appears if it exists.
In such a region, 
$\Kp$ indicates a high temperature (small $\Kp$),
and the relaxations
$\left[\big\langle \rme^{\rmi\phi_i (t+\tw) -\rmi\phi_i (\tw)}
\big\rangle_K^\Kp \right]_\rmc$ and 
$\left[\big\langle \cH (t)
\big\rangle_K^\Kp \right]_\rmc$ for the process I
describe a ZFC process 
(cooled immediately from a high temperature with $\Kp$).
On the other hand, 
$\left[\big\langle \rme^{\rmi\phi_i (t+\tw)-\rmi\phi_i(\tw)} 
\big\rangle_K^\bF \right]_\rmc$ and 
$\left[\big\langle \cH (t) 
\big\rangle_K^\bF \right]_\rmc$ for the process II
are regarded as the relaxations 
in a FQ process
(demagnetization from the strong field limit).
Thus, relations (\ref{AR}) and (\ref{jHH}) imply the equivalence of 
the dynamics especially the aging phenomenon in
the ZFC and the FQ processes.
The same relation was pointed out 
in real SG materials \cite{NordEA87,LeflEA94}.
At a glance, this relation expresses a trivial fact,
since both initial states are located in the PM phase and 
both systems relax into the SG phase.
However, the equivalence at any waiting time, 
which means the equivalence of the dynamical structure
at any stage of relaxation, is non-trivial. 

In the case of sufficiently weak randomness (large $\Kp$),
the initial states of both nonequilibrium processes represent
almost the same spin state (complete FM state).
Then, equations (\ref{AR}) and (\ref{jHH}) show a proper phenomenon.

\subsection{Dynamics on $K=\Kp$}
Next, we examine the dynamics on the line $K=\Kp$,
where the dynamical average $\langle\cdots\rangle_K^\Kp$
becomes the average in the equilibrium process
$\langle\cdots\rangle_\Kp^\eq$.
{}From eqs.\ (\ref{ACeq}) and (\ref{AR}), we derive
\begin{equation}
\left[\big\langle \rme^{\rmi\phi_i (t+\tw)-\rmi\phi_i(\tw)} 
\big\rangle_\Kp^\bF \right]_\rmc ~=~
\left[\big\langle \rme^{\rmi\phi_i (t) -\rmi\phi_i (0)}
\big\rangle_\Kp^\eq \right]_\rmc.
\label{eq:AReq}
\end{equation}
If one sets $\tw=0$, equation (\ref{eq:AReq}) yields 
\begin{equation}
\left[\big\langle \rme^{\rmi\phi_i (t)} 
\big\rangle_\Kp^\bF \right]_\rmc ~=~
\left[\big\langle \rme^{\rmi\phi_i (t) -\rmi\phi_i (0)}
\big\rangle_\Kp^\eq \right]_\rmc,
\label{eqMQ}
\end{equation}
which implies the equivalence of FM and SG orderings 
(if one takes the limit of $t\to\infty$).
Since, in the SG phase, the SG order remains finite whereas 
the FM order disappears, 
eq.\ (\ref{eqMQ}) indicates the absence of the SG phase 
on the line $K=\Kp$
as in the analysis of the static orderings (\ref{eq:mq}) \cite{OzekiN93}.

Equations (\ref{eq:AReq}) and (\ref{eqMQ}) also imply the equivalence 
of the equilibrium and nonequilibrium relaxations.
The internal energy of the initial state $\bF$ is 
identical with that on $K=\Kp$,
which is equal to the equilibrium energy on $K=\Kp$ 
(\ref{eq:E}) \cite{OzekiN93}.
It is easy to see that, the rhs of eq.\ (\ref{jHH}) 
is independent of time $t$ if $K=\Kp$;
\begin{equation}
\left[\big\langle \cH(t)\big\rangle_\Kp^\bF \right]_\rmc ~=~
-N_\rmB J\, {I_1 (\Kp ) \over I_0 (\Kp ) } .
\end{equation}
Since the equilibrium energy is an increasing function of the temperature,
we expect that 
$\left[\big\langle \cH (t) 
\big\rangle_K^\bF \right]_\rmc$ 
is an increasing function of $t$ above $K=\Kp$ 
and a decreasing one below it.

As seen from eqs.\ (\ref{eq:AReq}) and (\ref{eqMQ}),
the auto-correlation function in the FQ process is independent of the
waiting time, which suggests the absence of aging on $K=\Kp$.
In the region below the MCP on $K=\Kp$, where the spontaneous
ordering appears,
it is not clear that the FQ process is appropriate 
for the observation of the aging,
since the initial all-aligned state is not so different 
from the final equilibrium state with broken symmetry.
While the rotational symmetry is initially broken in the FQ process,
the dynamical behavior is equivalent to that in the equilibrium 
(symmetric) process --- see eq.\ (\ref{eq:AReq}).
Thus, we assume that the FQ process exhibits an aging feature 
even in such a symmetry-broken region.
\subsection{Mixed Phase and Griffiths Phase}

The aging has been considered an inherent property in the SG phase
from experimental \cite{LundEA83,NordEA87,VincHO92,LeflEA94}
and theoretical \cite{KoperH88,FisheH88,SibanH89,Rieger93,CugliK94} 
view points.
Since the aging phenomenon
is a typical observation of the complex phase space for slow dynamics,
it could occur in other complex phases.
In such a phase, the aging would also be inherent, which means that 
it occurs whole of the phase when it is observed in some parts of the phase.

Following the results of the SK model and experimental observations 
\cite{BindeY86} for SG phenomenon,
one may consider a possibility of 
the mixed phase between the FM and SG phases 
even in the gauge glass systems.
It is natural to consider that the aging occurs in the mixed phase, 
since the SG feature in the mixed phase provides
typical slow dynamics which reveals the aging.
If the aging is inherent in the mixed phase,
this indicates that 
the line $K=\Kp$ does not enter the mixed phase.
This restricts the topology of the phase diagram:
it must appear below $K=\Kp$ even if it exists.

In the Ising SG systems,
it has also been pointed out that there exists a dynamically singular phase
called the Griffiths phase
\cite{Griffi69}
between the critical temperature
of the pure system and the phase boundary of low temperature phase 
(the FM or the SG)
\cite{RandSP85,Ogiels85,TakanM95,KomoHT95}
--- see Fig.\ \ref{PhaseD}.
The possibility of the Griffiths phase has not yet been 
discussed much in the gauge glass systems.
Figure \ref{PhaseD} is proposed just following an analogy 
with the dilute ferromagnet \cite{Griffi69}.
Concerning to the Griffiths phase,
two cases can be considered to understand the present result
if the aging is an inherent property:
(a) {\it There is no aging whole of the Griffiths phase},
if the line $K=\Kp$ intersects the Griffiths phase as in Fig.\ \ref{PhaseD}.
(b) {\it There is no Griffiths phase at least around $K=\Kp$.}
In the former case, even if slow dynamics is observed in the Griffiths phase,
it is quite different from that in the SG phase \cite{DerriW87}.
The latter case allows the existence of the Griffiths phase 
below $K=\Kp$.
Another possibility is that 
(c) {\it the aging is not inherent at least in the Griffiths phase.}
This means that the aging occurs in some parts of the phase.
Although it is not obvious which is correct in the present framework,
the above result restricts the existence of the aging phenomenon
and the region of the Griffiths phase,
and supports future investigations of them.
\subsection{Multicritical Dynamics}
Let us consider the critical relaxation at the MCP.
At the FM critical point, in general,
the FM order parameter  
decays in a power-law asymptotically as 
\begin{equation}
m(t)
~\equiv~[\big\langle \rme^{\rmi\phi_i (t)} \big\rangle_K^\bF]_\rmc
~\sim~ t^{-\lambda_m}.
\label{eq:asymm}
\end{equation} 
Then, the dynamic scaling hypothesis \cite{Suzuki76}
\begin{equation}
m(t, \varepsilon, L)=L^{-\beta/\nu} 
\bar{m}(L^{1/\nu}\varepsilon, L^{-z} t)
~~~~~\varepsilon =(K_\rmc-K)/K_\rmc
\label{eq:scmtl}
\end{equation}
reveals the relation of exponents
\begin{equation}
\lambda_m={\beta\over z\nu}.
\label{eq:expmf}
\end{equation}
On the other hand,
the equilibrium auto-correlation function decays in a power-law as
\begin{equation}
q(t)
~\equiv~
\left[\big\langle \rme^{\rmi\phi_i (t) -\rmi\phi_i (0)}
\big\rangle_K^\eq \right]_\rmc
~\sim~ t^{-\lambda_q}
\label{eq:asymq}
\end{equation} 
at both FM and SG critical points.
At the FM critical point, the exponent $\lambda_q$ is associated with
the FM criticality;
a similar dynamic scaling reveals 
\begin{equation}
\lambda_q={2\beta\over z\nu}.
\label{eq:expqf}
\end{equation}
Thus, $\lambda_m$ and $\lambda_q$ are twice different 
in the FM case.
At the SG critical point, it is associated with the SG criticality;
the dynamic scaling hypothesis for the SG ordering
\begin{equation}
q(t, \varepsilon, L)=L^{-\tilde{\beta}/\tilde{\nu}} 
\bar{q}(L^{1/\tilde{\nu}}\varepsilon, L^{-\tilde{z}} t),
\label{eq:scqtl}
\end{equation}
where tilde exponents are those for SG ordering, yields
\begin{equation}
\lambda_q={\tilde{\beta}\over\tilde{z}\tilde{\nu}}.
\label{eq:expqg}
\end{equation}

The scaling field is not unique in the random case.
One can define another scaling form instead of (\ref{eq:scmtl})
with another scaling field such as $(\Kp_\rmc-\Kp)/\Kp_\rmc$.
At the MCP, the scaling field $\varepsilon$ can be chosen
both as $\varepsilon =(\Kp_\rmc-\Kp)/\Kp_\rmc$ and
as $\varepsilon =(K_\rmc-K)/K_\rmc$ 
since they are same order around the MCP on $K=\Kp$.
Note that critical exponents $\beta$ and $\nu$ depend on the way
to approach the critical point.
Thus, they are defined along $K=\Kp$
in eqs.\ (\ref{eq:scmtl}) and (\ref{eq:expmf}) for the MCP case.
On the other hand, 
the ratio $\beta /\nu$ and the exponent $z$ are independent of 
the scaling field but depend on the critical point itself.
On $K=\Kp$, we have derived the relation (\ref{eqMQ}).
This gives
\begin{equation}
L^{-\beta/\nu} \bar{m}(L^{1/\nu}\varepsilon, L^{-z} t)
=L^{-\tilde{\beta}/\tilde{\nu}} 
\bar{q}(L^{1/\tilde{\nu}}\varepsilon, L^{-\tilde{z}} t),
\label{eq:lmelq}
\end{equation}
for any $L$, $t$ and $\varepsilon$ ($K$ or $\Kp$)
providing
\begin{eqnarray}
\lambda_m&=& \lambda_q ,
\label{eq:dmedq}
\\
z&=&\tilde{z},
\label{eq:zmezq}
\\
\beta/\nu &=&\tilde{\beta}/\tilde{\nu}.
\label{eq:bmebq}
\end{eqnarray}
Note that the line $K=\Kp$ does not enter the SG phase but just touch
it at the MCP 
\cite{OzekiN93}.
The dynamical exponent $\tilde{z}$ and the ratio 
$\tilde{\beta}/\tilde{\nu}$
are independent of the way to approach the criticality,
and characterized by the criticality itself.
Therefore, exponents $\tilde{z}$ 
and $\tilde{\beta}/\tilde{\nu}$ are associated 
with the SG ordering at the MCP.
Equations (\ref{eq:dmedq})-(\ref{eq:bmebq}) 
relate exponents of FM ordering and SG ordering at the MCP.
The relations for FM dynamic exponents, (\ref{eq:expmf}) and (\ref{eq:expqf}),
would keep along the phase boundary
from the pure case up to the MCP in Fig.\ \ref{PhaseD},
and so does the relation for SG ones (\ref{eq:expqg}) 
from $\Kp=0$ up to the MCP;
it is not clear whether the values of exponents are universal or not.

\subsection{Absence of Re-entrance}
Here, we discuss the asymptotic behavior of
the auto-correlation function for the process I with $\tw=0$, 
$\left[\big\langle \rme^{\rmi\phi_i (t) -\rmi\phi_i (0)}
\big\rangle_K^\Kp \right]_\rmc$, 
which satisfies
\begin{equation}
\left[\big\langle \rme^{\rmi\phi_i (t)} 
\big\rangle_\Kp^\bF \right]_\rmc ~=~
\left[\big\langle \rme^{\rmi\phi_i (t) -\rmi\phi_i (0)}
\big\rangle_K^\Kp \right]_\rmc,
\label{neMQ}
\end{equation}
to show the absence of re-entrant transition,
{\it i.e.} the verticality of the FM boundary.
We present the following two propositions for it.
Since the long-time limit of this function is 
expected to behave a kind of order parameter, 
(i) {\it the asymptotic behavior should be
prescribed by both phases where the initial point $(\Kp ,\Kp )$ 
and the final point $({K,\Kp} )$ locate.}
In Appendix \ref{sec:MM},
we mention that the average of gauge invariant function
can not detect the FM boundary.
Because of the gauge invariance (\ref{giACeq}),
(ii) {\it the asymptotic behavior is independent of 
whether $(\Kp ,\Kp )$ and $({K,\Kp} )$ are FM or not.}
On this point of view, 
the behavior of 
$\left[\big\langle \rme^{\rmi\phi_i (t)-\rmi\phi_i
(0)}\big\rangle_K^\Kp\right]_\rmc$
can change only at which $(\Kp,\Kp)$ or $(K,\Kp)$ intersects the PM
boundary.
On the other hand, it approaches an FM order parameter
because of eq.\ (\ref{neMQ}), implying that the asymptotic behavior 
changes depending on if $({K,\Kp} )$ is FM or not.
At a glance, this contradicts with (ii) on the FM boundary below the PM
boundary.
However, they are not in conflict with each other if 
the line $K=\Kp$ intersects the MCP
and the FM phase boundary is vertical in the temperature below it.
In this case, when $(K,\Kp)$ locates on the FM boundary, $(\Kp,\Kp)$
locates on the PM boundary simultaneously ({\it i.e.} the MCP).

In two dimensions, the same argument can be made if one considers
the asymptotic decay of relaxation functions as an order parameter
for the KT phase instead of the long time limit of them.
The order parameter relaxation $\left[\big\langle \rme^{\rmi\phi_i (t)}
\big\rangle_K^\bF \right]_\rmc$
decays exponentially in the PM phase while it decays algebraically in
the KT phase. 
Thus, the above argument is recognized as the statement 
that the KT boundary is vertical below the MCP.

\section{Deterministic Dynamics}
\label{sec:HD}
In section \ref{sec:SD}, we treat the stochastic dynamics 
obeying the master equation with continuous time (\ref{eqMaster}).
The results can also be applied to the discrete time
interval case, which is used in Monte Carlo simulations. 
Since the spin variable is continuous in the gauge glass model, 
one can define equations of motion.
In this section, we discuss the case of the deterministic dynamics
like in molecular dynamics simulations.

It is noted that the essential features for a physical quantity $Q$
to derive relations like 
(\ref{AR}) and (\ref{jHH}) are the gauge invariance
\begin{equation}
V_\bpsi~
\langle Q\rangle_K^\Kp ~=~
\langle Q\rangle_K^\Kp
\label{giQ}
\end{equation}
for the process I as eq.\ (\ref{giAC1}), and the transformation rule
\begin{equation}
V_\bpsi~
\langle Q\rangle_K^\bF ~=~
\langle Q\rangle_K^\bpsi
\label{gtQ}
\end{equation}
for the process II as eq.\ (\ref{gtAC2}).
By use of eqs.\ (\ref{rAVgt}) and (\ref{rAVgi}), 
they reveal the relation
\begin{equation}
[\langle Q\rangle_K^\bF]_\rmc ~=~
[\langle Q\rangle_K^\Kp]_\rmc
\label{ARQ}
\end{equation}
irrespective of the detail of dynamics.
\subsection{Microcanonical Dynamics}
Let us introduce a conjugate momentum $L_i$ for each spin variable
$\phi_i$ and consider the total Hamiltonian 
$\cH_\total =
J\wcH_\total (\bL,\bphi,\bomega)$
including the kinetic
energy;
\begin{equation}
\wcH_\total (\bL,\bphi,\bomega) ~=~
-\sum_{\langle ij \rangle}\, \cos(\phi_i -\phi_j +\omega_{ij} )
+ {1\over 2}\sum_i L_i^2 .
\label{defHtot}
\end{equation}
The thermal distribution for this system is given by
\begin{equation}
{
\exp\left(-K\wcH_\total (\bL,\bphi,\bomega)\right)
\over
Z_\total (K;\bomega )}
~=~
\mu_\eq (\bL;K)~\rho_\eq (\bphi;K,\bomega),
\label{rhotot}
\end{equation}
where the partition function of $\cH_\total$ is defined by
\begin{equation}
Z_\total(K;\bomega )~=~
\int \rmd\bL\,\rmd\bphi~
\exp\left(-K\wcH_\total (\bL,\bphi,\bomega)\right).
\end{equation}
The integration of $\rmd\bL$ denotes the multiple integrations for
$N$ momentum variables.
For the spin variable, the thermal distribution of 
$\cH_\total$ is identical with that in the static
case (\ref{rhoeq}).
The function
$\mu_\eq (\bL;K)$ is the thermal distribution for momentum variables;
\begin{equation}
\mu_\eq (\bL;K)\equiv
\left(K\over2\pi\right)^{N/2} 
\exp(-{K\over 2}\sum_i L_i^2).
\end{equation}
The equations of motion for $\phi_i$ and $L_i$ are given by 
the canonical equations of Hamilton;
\begin{subequations}
\begin{align}
{\partial L_i\over \partial t} 
&=-{\partial \over\partial\phi_i}\cH_\total
=-J\sum_{(j\in i)} \sin(\phi_i-\phi_j+\omega_{ij}),
\\ 
{\partial \phi_i\over \partial t} 
&=~~\,{\partial \over\partial L_i}\cH_\total
=JL_i,
\end{align}
\label{kinEQ}
\end{subequations}
where the summation of $(j\in i)$ takes for all sites
coupled with the $i$ site. 

Note that the conjugate momentum $L_i$, differential operators
$\displaystyle{\partial \over\partial L_i}$ and 
$\displaystyle{\partial \over\partial\phi_i}$
are invariant under gauge transformations $U_\bpsi$ and $V_\bpsi$.
Thus, 
the canonical equations (\ref{kinEQ}) are also invariant 
under the gauge transformation $U_\bpsi V_\bpsi$.
This means that if $(\bar{\bL}(t),\bar{\bphi}(t))$ 
represents the solution for a bond configuration $\bomega$
with an initial condition $(\bL_0,\bphi_0)$, 
the solution for the bond configuration $V_\bpsi\,\bomega$
with the initial condition $(\bL_0,\bphi_0-\bpsi)$ is given by
$(\bar{\bL}(t),\bar{\bphi}(t)-\bpsi)$. 
This implies the transformation rules
\begin{subequations}
\begin{align}
U'_\bpsi\,V_\bpsi~ 
\bar{\phi}_i (t;\bL_0,\bphi_0,\bomega) &=
\bar{\phi}_i (t;\bL_0,\bphi_0,\bomega) -\psi_i,
\\
U'_\bpsi\,V_\bpsi~ 
\bar{L}_i (t;\bL_0,\bphi_0,\bomega) &=
\bar{L}_i (t;\bL_0,\bphi_0,\bomega), 
\end{align}
\label{UVkin}
\end{subequations}
where $U'_\bpsi$ is the transformation operated to $\bphi_0$.
It is remarked that once the transformation for $\bphi_0$ is operated, it
affects $\bphi(t)$ at any time, since the solution is continuous in time.
Thus, the operation $U'_\bpsi\,V_\bpsi$ in eqs.\ (\ref{UVkin})
can be recognized as $U_\bpsi\,U'_\bpsi\,V_\bpsi$
in the stochastic dynamics.

Let us define dynamical averages like eqs.\ (\ref{AC1})-(\ref{ACpsi}).
Since the dynamics is deterministic,
the dynamical average is performed by the average
over the initial condition $(\bL_0,\bphi_0)$.
When the initial distribution is in equilibrium,
the temperature is unchanged since the total energy is conserved.
The equilibrium dynamics with temperature $K$, 
$\big\langle \cdots \big\rangle_K^\eq =
\big\langle \cdots \big\rangle_K^K$, is defined as
\begin{equation}
\big\langle \cdots \big\rangle_K^\eq 
=
\int \rmd\bL_0\,\rmd\bphi_0~
\mu_\eq (\bL_0;K)~\rho_\eq (\bphi_0;K,\bomega)~
\cdots.
\label{aveEQ}
\end{equation}
Although the dynamics contains no dissipative factor
in this formulation, 
we assume that the system shows a relaxation to an orbit which 
traces the thermal distribution
when one prepares an appropriate nonequilibrium initial distribution
$\rho_0(\bL_0,\bphi_0)$.
This is relating to the ergodic theorem.
There are several ways to determine $\rho_0(\bL_0,\bphi_0)$.
One simplest way is to fix the total energy to the equilibrium value
at the final temperature.  
In the process I,
$\langle\cdots\rangle_K^{\Kp}$,
the initial distribution for spin variables is the thermal one with 
temperature $\Kp$.
The total energy is identical with the equilibrium value at
temperature $K$ providing 
\begin{equation}
\rho_\rmI(\bL_0,\bphi_0;K,\Kp,\bomega)=
{\rho_\eq (\bphi_0;\Kp,\bomega)\over
\Omega_\rmI(K,\Kp,\bomega)}~
\delta
\left(\wcH_\total (\bL_0,\bphi_0,\bomega)-\langle \wcH_\total\rangle_K
\right).
\label{idM1}
\end{equation}
where $\langle \wcH_\total\rangle_K = N/2K+\langle \wcH\rangle_K$, and
\begin{equation}
\Omega_\rmI(K,\Kp,\bomega) ~=~
\int\rmd\bL_0\,\rmd\bphi_0~
\rho_\eq (\bphi_0;\Kp,\bomega)~
\delta
\left(\wcH_\total (\bL_0,\bphi_0,\bomega)-\langle \wcH_\total\rangle_K
\right)
\end{equation}
is the volume of the phase space in the process I.
When $K=\Kp$, the dynamical average with $\rho_\rmI$ becomes identical with
(\ref{aveEQ}), because of the equivalence of the canonical and
microcanonical ensembles in large enough systems.
Similarly, the initial distribution for the process II' is given by
\begin{equation}
\rho_{\bpsi}(\bL_0,\bphi_0;K,\bomega)~=~
{\delta (\bphi_0-\bpsi)\over
\Omega_\bpsi (K,\bomega)}~
\delta
\left(\wcH_\total (\bL_0,\bphi_0,\bomega)-\langle \wcH_\total\rangle_K
\right).
\label{idM2}
\end{equation}
where
\begin{equation}
\Omega_\bpsi (K,\bomega) ~=~
\int\rmd\bL_0\,\rmd\bphi_0~
\delta (\bphi_0-\bpsi)~
\delta
\left(\wcH_\total (\bL_0,\bphi_0,\bomega)-\langle \wcH_\total\rangle_K
\right)
\end{equation}
is the volume of the phase space in this process.
Using these initial distributions,
we define the correlation functions for processes I and II;
\begin{eqnarray}
\label{defDKL}
\big\langle \rme^{\rmi\phi_i (t+\tw) -\rmi\phi_i (\tw)} 
\big\rangle_K^D 
&=&\int \rmd\bL_0\,\rmd\bphi_0~
\rho_\rmI (\bL_0,\bphi_0;K,\Kp,\bomega)~ 
\rme^{\rmi\bar{\phi}_i (t+\tw;\bL_0,\bphi_0,\bomega)
-\rmi\bar{\phi}_i (\tw;\bL_0,\bphi_0,\bomega)},
\nonumber\\
\\
\label{defFKL}
\big\langle \rme^{\rmi\phi_i (t+\tw)-\rmi\phi_i (\tw) } 
\big\rangle_K^\bF &=&
\int \rmd\bL_0\,\rmd\bphi_0~ 
\rho_\bF (\bL_0,\bphi_0;K,\bomega)~ 
\rme^{\rmi\bar{\phi}_i (t+\tw;\bL_0,\bphi_0,\bomega)
-\rmi\bar{\phi}_i (\tw;\bL_0,\bphi_0,\bomega)}.
\end{eqnarray}
The dynamical averages for the energy are also defined in a similar way.

Noting the invariance (\ref{girhoeq})-(\ref{giAH}), 
the function $\rho_\rmI (\bL_0, \bphi_0;K,\Kp,\bomega)$ is invariant
under the gauge transformation $U'_\bpsi V_\bpsi$;
\begin{equation}
U'_\bpsi V_\bpsi~\rho_\rmI (\bL_0,\bphi_0;K,\Kp,\bomega)~=~ 
\rho_\rmI (\bL_0,\bphi_0;K,\Kp,\bomega).
\label{girho0}
\end{equation}
This reveals the gauge invariance of
$\big\langle \rme^{\rmi\phi_i (t+\tw) -\rmi\phi_i (\tw)} 
\big\rangle_K^D$ for the process I as in the stochastic dynamics case 
(\ref{giAC1}).
On the other hand, the transformation for $\rho_\bF$ gives
\begin{equation}
U'_\bpsi V_\bpsi~\rho_{\bF}(\bL_0,\bphi_0;K,\bomega)~=~
\rho_{\bpsi}(\bL_0,\bphi_0;K,\bomega)
\end{equation}
which reveals the transformation rule for the process II,
\begin{eqnarray}
V_\bpsi~\big\langle \rme^{\rmi\phi_i (t+\tw)-\rmi\phi_i (\tw) } 
\big\rangle_K^\bF &=&
\int \rmd\bL_0\,\rmd\bphi_0~ 
\rho_\bpsi (\bL_0,\bphi_0;K,\bomega)~ 
\rme^{\rmi\bar{\phi}_i (t+\tw;\bL_0,\bphi_0,\bomega)
-\rmi\bar{\phi}_i (\tw;\bL_0,\bphi_0,\bomega)}
\nonumber\\
&=&
\big\langle \rme^{\rmi\phi_i (t+\tw)-\rmi\phi_i (\tw) } 
\big\rangle_K^\bpsi 
\label{gtHF}
\end{eqnarray}
Thus, form eqs.\ (\ref{giQ})-(\ref{ARQ}),
equation (\ref{gtHF}) 
yields eq.\ (\ref{cavAC2}) 
providing the aging relation (\ref{AR}).
It is easy to see that eq.\ (\ref{jHH}) also holds.
\subsection{Liouville Equation}
To compare the formulations in stochastic and deterministic 
dynamics discussed above,
it is instructive to consider the classical Liouville equation
which describes the time evolution of distribution function $\rho_t$ 
under the canonical equations (\ref{kinEQ});
\begin{equation}
{\rmd\rho_t\over\rmd t} =
\sum_i\left\{
{\partial\cH_\total\over\partial L_i}
{\partial\rho_t\over\partial \phi_i}
-
{\partial\cH_\total\over\partial \phi_i}
{\partial\rho_t\over\partial L_i}
\right\}.
\label{Liouville}
\end{equation}
If we define the solution of eq.\ (\ref{Liouville}) for the initial condition
$\rho =\delta(\bL-\bL_0)\delta(\bphi-\bphi_0)$
as $G(t,\bL,\bphi;\bL_0,\bphi_0,\bomega)$, this is nothing but a delta
function
\begin{equation}
G(t,\bL,\bphi;\bL_0,\bphi_0,\bomega)=
\delta\left(\bL-\bar{\bL}(t;\bL_0,\bphi_0,\bomega)\right)~
\delta\left(\bphi-\bar{\bphi}(t;\bL_0,\bphi_0,\bomega)\right),
\label{rhodelta}
\end{equation}
since the orbit is deterministic.
The function $G$ is called the Green function, and 
the solution for any initial distribution $\rho_0(\bL,\bphi)$
is given by
\begin{equation}
\rho_t(\bL,\bphi;\bomega,[\rho_0])~=~
\int \rmd\bL_0\rmd\bphi_0
~G(t,\bL,\bphi;\bL_0,\bphi_0,\bomega)
~\rho_0(\bL_0,\bphi_0),
\label{defrhot}
\end{equation} 
where $\rho_t$ is a functional of the initial distribution  
indicated in the arguments as $[\rho_0]$.

The Green function $G$ plays a similar role with 
$\left\langle\bphi \left|\rme^{t\bW}\right|\bphi'\right\rangle$ 
in the stochastic dynamics.
It is easy to see that eqs.\ (\ref{UVkin}) and
(\ref{rhodelta}) yield the invariance, 
\begin{equation}
U_\bpsi\,U'_\bpsi\,V_\bpsi~ G(t,\bL,\bphi;\bL_0,\bphi_0;\bomega) ~=~ 
G(t,\bL,\bphi;\bL_0,\bphi_0;\bomega), 
\label{giG}
\end{equation}
which is the same relation as eq.\ (\ref{giETW}).
Since the differential operators and $\cH_\total$ is 
invariant under $U_\bpsi V_\bpsi$,
when the function $\rho_t$ is the solution of eq.\ (\ref{Liouville}) 
with an initial condition $\rho_0$, 
the function $U_\bpsi\,V_\bpsi\,\rho_t$ is also the
solution with the initial condition $U_\bpsi\,\rho_0$:
It is easily confirmed from eq.\ (\ref{defrhot}) that
\begin{equation}
U_\bpsi\,V_\bpsi~ 
\rho_t(\bL,\bphi;\bomega,[\rho_0])~=~
\rho_t(\bL,\bphi;\bomega,[U_\bpsi\,\rho_0]).
\end{equation}
Using the definitions above with the Green function,
the auto-correlation functions are expressed as
\begin{eqnarray}
\big\langle \rme^{\rmi\phi_i (t+\tw) -\rmi\phi_i (\tw)} 
\big\rangle_K^\Kp ~=~
\int \rmd\bL\,\rmd\bphi\,\rmd\bL'\,\rmd\bphi'\,\rmd\bL_0\,\rmd\bphi_0
&&\\
~G(t,\bL,\bphi;\bL',\bphi',\bomega)
~G(\tw,\bL',\bphi';\bL_0,\bphi_0,\bomega)
&&
\rho_\rmI (\bL_0,\bphi_0;K,\Kp,\bomega)~ 
~\rme^{\rmi\phi_i -\rmi\phi'_i},
\nonumber
\end{eqnarray}
for the process I, and
\begin{eqnarray}
\big\langle \rme^{\rmi\phi_i (t+\tw) -\rmi\phi_i (\tw)} 
\big\rangle_K^\bF ~=~ 
\int \rmd\bL\,\rmd\bphi\,\rmd\bL'\,\rmd\bphi'\,\rmd\bL_0\,\rmd\bphi_0
&&\\
~G(t,\bL,\bphi;\bL',\bphi',\bomega)
~G(\tw,\bL',\bphi';\bL_0,\bphi_0,\bomega)
&&
\rho_\bF (\bL_0,\bphi_0;K,\bomega)~ 
~\rme^{\rmi\phi_i -\rmi\phi'_i}.
\nonumber
\end{eqnarray}
for the process II.
They are equivalent with eqs.\ (\ref{defDKL}) and (\ref{defFKL}).
{}From invariance (\ref{giG}) and (\ref{girho0})
with eq.\ (\ref{iiPU}), one finds that
$\big\langle \rme^{\rmi\phi_i (t+\tw) -\rmi\phi_i (\tw)} 
\big\rangle_K^D$ is gauge invariant and the other one obeys
\begin{eqnarray}
\label{gtHFL}
V_\bpsi~\big\langle \rme^{\rmi\phi_i (t+\tw) -\rmi\phi_i (\tw)} 
\big\rangle_K^\bF ~=~ 
\int \rmd\bL\,\rmd\bphi\,\rmd\bL'\,\rmd\bphi'\,\rmd\bL_0\,\rmd\bphi_0
&&\\
~G(t,\bL,\bphi;\bL',\bphi',\bomega)
~G(\tw,\bL',\bphi';\bL_0,\bphi_0,\bomega)
&&
\rho_\bpsi (\bL_0,\bphi_0;K,\bomega)~ 
~\rme^{\rmi\phi_i -\rmi\phi'_i}.
\nonumber
\end{eqnarray}
Thus, form eqs.\ (\ref{giQ})-(\ref{ARQ}), 
this yields eq.\ (\ref{cavAC2}) 
providing the aging relation (\ref{AR}).

\subsection{Canonical Initial Distribution}
Another way to define the initial distributions instead of 
(\ref{idM1}) and (\ref{idM2}) is given by the canonical ensemble.
In both processes, the initial distribution is determined so that
the total energy distributes in terms of the thermal distribution with
temperature $K$;
\begin{equation}
\int \rmd\bL_0\,\rmd\bphi_0~
\rho_0 (\bL_0,\bphi_0)~
\delta\left(E-\cH_\total (\bL_0,\bphi_0,\bomega)\right) ~=~
{{\cal D}_\total (E;\bomega)~
\rme^{-KE/J} 
\over
Z_\total (K;\bomega)}.
\label{idC}
\end{equation}
where ${\cal D}_\total (E;\bomega)$ is the density of state;
\begin{equation}
{\cal D}_\total (E;\bomega)
=
\int \rmd\bL_0\,\rmd\bphi_0~
\delta(E-\cH_\total).
\end{equation}
For the process I, $\rho_0$ has a form of
\begin{equation}
\rho_\rmI(\bL_0,\bphi_0)~=~
\mu_\rmI (\bL_0,\bphi_0;K,\Kp,\bomega)~\rho_\eq (\bphi_0;\Kp,\bomega).
\label{defidC1}
\end{equation}
Equation (\ref{idC}) implicitly defines the function 
$\mu_\rmI (\bL_0,\bphi_0; K,\Kp,\bomega)$ depending on 
$K$, $\Kp$ and $\bomega$.
For the case $K=\Kp$, 
$\mu_\rmI=\mu_\eq (\bL_0; \Kp)$ satisfies eq.\ (\ref{idC}).
For the process II', $\rho_0$ has a form of
\begin{equation}
\rho_\bpsi (\bL_0,\bphi_0)~=~
\mu_\bpsi (\bL_0,\bphi_0;K,\bomega)~\delta(\bphi_0-\bpsi).
\end{equation}
Equation (\ref{idC}) implicitly defines the function 
$\mu_\bpsi (\bL_0,\bphi_0; K,\bomega)$ depending on 
$K$ and $\bomega$.

Noting the invariance (\ref{giH}), (\ref{giZ}) and (\ref{girhoeq}) with 
the invariant integral (\ref{iiPU}),
one finds that, if $\rho_0 (\bL_0,\bphi_0)$ satisfies eq.\
(\ref{idC}), so does $U'_\bpsi V_\bpsi~\rho_0 (\bL_0,\bphi_0)$.
Since $U'_\bpsi V_\bpsi~ \mu_\rmI$ instead of $\mu_\rmI$ satisfies
the same equation as (\ref{idC}) with (\ref{defidC1}),
the function $\rho_\rmI (\bL_0,\bphi_0)$ is invariant
under $U'_\bpsi V_\bpsi$.
This reveals the gauge invariance of
$\big\langle \rme^{\rmi\phi_i (t+\tw) -\rmi\phi_i (\tw)} 
\big\rangle_K^D$ for the process I.
On the other hand, if $\rho_\bF (\bL_0,\bphi_0)$ satisfies eq.\
(\ref{idC}) for the process II, so does
\begin{equation}
U'_\bpsi V_\bpsi~\rho_\bF (\bL_0,\bphi_0)~
=[U'_\bpsi V_\bpsi~\mu_\bF (\bL_0,\bphi_0;K,\bomega)]~\delta(\bphi_0-\bpsi).
\end{equation}
This means 
\begin{equation}
U'_\bpsi V_\bpsi~\rho_\bF (\bL_0,\bphi_0;K,\bomega)~
=\rho_\bpsi (\bL_0,\bphi_0;K,\bomega).
\end{equation}
Therefore one can derive the transformation rules like (\ref{gtHF})
providing (\ref{cavAC2}) and the aging relation (\ref{AR}) 
for the canonical case.

\subsection{Extended System}
In the above, we have defined relaxation processes 
in conservative dynamics with gauge symmetry. 
Here we consider another deterministic dynamics, which keeps the 
temperature constant \cite{Nose84,KogutS86}.
It is called the extended system dynamics.

In addition to the momentum variables, we introduce a single degree of
freedom, the daemon, which couples to the original system.
It will act as a heat bath for the original system
and convert the dynamics as an fixed-temperature one.
Let the variable $y$ be the daemon and $\ell$ the conjugate momentum of
it.
The Hamiltonian of the extended system 
$\cH_\extend =J\wcH_\extend  
(\widetilde{\bL},\widetilde{\bphi},\bomega )$ is defined by
\begin{equation}
\wcH_\extend (\widetilde{\bL},\widetilde{\bphi},\bomega )~=~
-\sum_{\langle ij \rangle}\, \cos(\phi_i -\phi_j +\omega_{ij} )
+ {1\over 2 y^2}\sum_i L_i^2 
+{\ell^2\over 2} +{(N+1)\over K}\ln y,
\label{defHext}
\end{equation}
where $\widetilde{\bphi}$ represents the set of spin variables $\bphi$
including the daemon $y$, and $\widetilde\bL$ represents the set of
momentum variables $\bL$ including $\ell$.
Note that the temperature $K$ is explicitly included in 
$\cH_\extend$.
The canonical equations are given by
\begin{subequations}
\begin{align}
{\partial L_i\over \partial t} 
&=
-J\sum_{(j\in i)} \sin(\phi_i-\phi_j+\omega_{ij}),
\\ 
{\partial \phi_i\over \partial t} 
&=
JL_i /y^2,
\\
{\partial \ell\over \partial t} 
&=
J\left(\sum_i L_i^2 /y^2 -(N+1)/K\right)/y,
\\
{\partial y\over \partial t} 
&=
J\ell.
\end{align}
\label{extEQ}
\end{subequations}
It is known that the microcanonical ensemble of $\cH_\extend$ with the
total energy $E$  becomes the canonical ensemble of
$\cH_\total$ with temperature $K$ irrespective of $E$ when
one traces out the daemon variables $y$ and $\ell$.
This means that the solution of the equations with an arbitrary initial
condition shows a relaxation to an orbit which traces the equilibrium 
distribution with temperature $K$,
if one assumes the ergodic theorem.

In this system, the dynamical averages are given by
\begin{equation}
\big\langle \cdots \big\rangle_K^\Kp 
~=~
\int \rmd\widetilde\bL_0\,\rmd\widetilde{\bphi}_0~
\mu (\bL_0,\ell,y)~\rho_\eq (\bphi_0;\Kp,\bomega)~
\cdots,
\end{equation}
for the process I and 
\begin{equation}
\big\langle \cdots \big\rangle_K^\bF 
=
\int \rmd\widetilde\bL_0\,\rmd\widetilde{\bphi}_0
~\mu (\bL_0,\ell,y)~\delta(\bphi_0 -\bF)~
\cdots,
\end{equation}
for the process II,
where the integrations of $\rmd\widetilde\bL_0$ and 
$\rmd\widetilde{\bphi}_0$ denote the multiple integrations for $N+1$
momentum variables including the daemon's $\ell$ and those for 
$N+1$ spin variables including $y$, respectively.
As stated above, the initial distribution function $\mu (\bL_0,\ell,y)$
for $\bL$, $\ell$ and $y$ can be chosen arbitrary.
We assume that $\mu (\bL_0,\ell,y)$ is independent of $\bomega$
so that it is gauge invariant.
It is natural to consider that the daemon variables $y$ and $\ell$
are invariant under the gauge transformation $U_\bpsi V_\bpsi$.
Thus, the canonical equations (\ref{extEQ})
are also invariant under the transformation,
which reveals the transformation rule like 
(\ref{UVkin});
\begin{equation}
U'_\bpsi\,V_\bpsi~ 
\bar{\phi}_i (t;\widetilde\bL_0,\widetilde{\bphi}_0,\bomega) =
\bar{\phi}_i (t;\widetilde\bL_0,\widetilde{\bphi}_0,\bomega) -\psi_i,
\label{UVphiext}
\end{equation}
Following the previous subsection, 
one can show the gauge invariance of 
$\big\langle \rme^{\rmi\phi_i (t+\tw) -\rmi\phi_i (\tw)} 
\big\rangle_K^D$ for the process I
and the transformation rule like (\ref{gtHF}) for the process II
providing (\ref{cavAC2}) and the aging-relation (\ref{AR}) 
for the extended system.
\section{Summary}
\label{sec:Sum}
We have applied the gauge transformation to dynamical systems 
of the gauge glass.
The stochastic dynamics with the master equation formalism
(\ref{eqMaster}) as well as the deterministic one with the canonical
equations of Hamilton are treated.
For the deterministic case, we consider standard equations of motion
with no dissipative factor (\ref{kinEQ}), in which 
one realizes relaxation processes to choose appropriate
initial distributions.
Both microcanonical and canonical ensembles are considered for 
the initial distribution.
Further, we consider the extended system dynamics (\ref{extEQ}) in which
the temperature is fixed.
In this dynamics, a single degree of freedom is introduced,
which is called the daemon and plays a role of a heat bath.
For all dynamics, we show the theory of gauge transformation
in a coherent manner.

The dynamical relations (\ref{AR}) and (\ref{jHH}) 
are derived exactly for all dynamics.
The equivalence of nonequilibrium relaxations in the SG phase
has been shown between the ZFC and FQ processes.
The waiting-time dependence, which is typical in the aging-phenomenon
coincides for ZFC and FQ processes.
On the line $K=\Kp$, we have found that the equilibrium relaxation 
coincides with the nonequilibrium one from the strong field limit.
This indicates the absence of aging providing the restrictions 
for the regime of possible mixed phase and Griffiths phase.
The exact relations for critical exponents are found at the MCP
on this line. 
Further, we have confirmed the verticality of the FM phase boundary 
or equivalently the absence of the re-entrant transition.

In contrast to the standard SG systems, 
not many studies on the dynamical properties have been performed for
gauge glass systems.
The present results would be helpful for future investigations 
including dynamics.
\acknowledgments{
This work is supported by a Grant-in-Aid for Scientific Research Program
(\#14540354) from the Ministry of Education, Culture, Sports, Science
and Technology of Japan.
}
\appendix
\section{Modified Model}
\label{sec:MM}
The modified model \cite{Kitata92,OzekiN93} is useful to consider 
the qualitative behaviors of physical quantities and 
the topology of phase diagram.
The Hamiltonian is identical with the original model.
The distribution of randomness, including a fixed constant $a$,
is modified from the original model ($a=0$ case),
\begin{equation}
P_a (\bomega ;\Kp )~\equiv~
P(\bomega ;\Kp +a)
{Y(\Kp +a)~Z(\Kp;\bomega)\over Y(\Kp)~Z(\Kp +a;\bomega)}.
\end{equation} 
The random average in the modified model with $a$ is denoted by
\begin{equation}
\{\cdots\}_\rmc^a ~\equiv~\int \rmd\bomega~
P_a (\bomega ;\Kp )~\cdots~.
\end{equation}
It is shown \cite{OzekiN93} 
that the average of any gauge invariant quantity is 
independent of $a$, {\it i.e.}
\begin{equation}
\left\{Q(\bomega)\right\}_\rmc^a~=~
\left[Q(\bomega)\right]_\rmc.
\end{equation}

In this appendix, we summarize the properties of the modified model
obtained previously.
Analyzing the FM and SG order parameters defined from 
the static FM and SG correlation functions,
\begin{eqnarray}
\label{statM}
m_a (K,\Kp )^2~&\equiv&~\lim_{R_{ij}\to\infty}
\Big|\left\{\big\langle \rme^{\rmi\phi_i -\rmi\phi_j}
\big\rangle_K \right\}_\rmc^a \Big| \\
\label{statQ}
q_a (K,\Kp )^2~&\equiv&~\lim_{R_{ij}\to\infty}
\Big\{\big|\big\langle \rme^{\rmi\phi_i -\rmi\phi_j}
\big\rangle_K \big|^2\Big\}_\rmc^a,
\end{eqnarray}
the following properties have been found exactly for the model with $a$:
\begin{itemize}
\item
The PM boundary, at which the edge of  $q_a =0$ locates, 
is unchanged with $a$.
\item
The ordered phase on $K=\Kp +a$ must be the FM phase,
since $m_a=q_a$ holds on it.
\item
The line $K=\Kp +a$ is likely to intersects the MCP.
\item
The FM boundary below $K=\Kp +a$ is vertical or re-entrant.
Further, the verticality can be shown 
if the ordered phase between the line $K=\Kp +a$ and the non-random case
($\Kp =\infty$) is always FM for the modified model with any $a$;
this assumption is quite plausible 
since both sides exhibit only PM-FM transition. 
\end{itemize}
These properties are identical with those for the original model
($a=0$).
In two dimensions,
one can treat KT transitions in the same way 
by analyzing the correlation length for FM and SG orderings,
\begin{eqnarray}
\label{statxi}
\xi_a (K,\Kp )~&\equiv&~\lim_{R_{ij}\to\infty}
\left|
{\partial\over\partial R_{ij}}
\ln\Big|\left\{\big\langle \rme^{\rmi\phi_i -\rmi\phi_j}
\big\rangle_K \right\}_\rmc^a \Big| \right|^{-1}\\
\tilde\xi_a (K,\Kp )~&\equiv&~\lim_{R_{ij}\to\infty}
\left|
{\partial\over\partial R_{ij}}
\ln\Big\{\big|\big\langle \rme^{\rmi\phi_i -\rmi\phi_j}
\big\rangle_K \big|^2\Big\}_\rmc^a\right|^{-1},
\end{eqnarray}
instead of order parameters (\ref{statM}) and (\ref{statQ}).

Since the FM correlation function as well as 
the FM and KT order parameters eqs.\ (\ref{statM}) and (\ref{statxi})
are not gauge invariant, 
the FM (KT) boundary changes with $a$:
At least on $K=\Kp+a$, the FM (KT) regime changes with $a$.
Thus, the qualitative behavior of averaged gauge invariant quantities 
at $(K,\Kp )$ should not be influenced by the fact 
whether $(K,\Kp )$ is FM (KT) or not.
In other words, any gauge invariant quantity can not be 
an order parameter for the FM (KT) phase in modified models including 
the original one ($a=0$).

\end{document}